\documentclass[aps,floatfix,preprint,prl,superscriptaddress]{revtex4}
\usepackage[utf8]{inputenc}
\usepackage{amsmath,amssymb,amsfonts,amsthm,array,dcolumn,nicefrac,units}
\usepackage{graphicx}
\usepackage{sans}
\usepackage{units}
\usepackage{color}
\usepackage[normalem]{ulem}
\usepackage{units}

\begin{document}
\definecolor{orange}{rgb}{1,0.5,0}

\title{Anomalously large $g$-factor of single atoms adsorbed on a metal substrate}
\author{B. Chilian}
\affiliation{Institute of Applied Physics, Hamburg University, Jungiusstrasse 11, D-20355 Hamburg, Germany}
\author{A. A. Khajetoorians}
\email{akhajeto@physnet.uni-hamburg.de}
\affiliation{Institute of Applied Physics, Hamburg University, Jungiusstrasse 11, D-20355 Hamburg, Germany}
\author{S. Lounis}
\affiliation{Forschungszentrum J\"{u}lich, Peter Gr\"unberg Institut and Institute for Advanced Simulation, 52425 J\"{u}lich, Germany}
\author{A. T. Costa}
\affiliation{Instituto de F\'isica, Universidade Fedeal Fluminense, 24210-340 Niter\'oi, RJ, Brazil}
\author{D. L. Mills}
\affiliation{Department of Physics and Astronomy, University of California Irvine, California, 92697 USA}
\author{J. Wiebe}
\email[corresponding author\\Email address: ]{jwiebe@physnet.uni-hamburg.de}
\affiliation{Institute of Applied Physics, Hamburg University, Jungiusstrasse 11, D-20355 Hamburg, Germany}
\author{R. Wiesendanger}
\affiliation{Institute of Applied Physics, Hamburg University, Jungiusstrasse 11, D-20355 Hamburg, Germany}

\begin{abstract}
We have performed inelastic scanning tunneling spectroscopy	(ISTS)
	on individual Fe atoms adsorbed on a Ag(111) surface. ISTS reveals a magnetization excitation with a lifetime of about $\unit[400]{fsec}$ which decreases linearly
	upon application of a magnetic field.  Astoundingly, we find that the $g$-factor, which characterizes the shift in energy of the excitation in a magnetic field, is $g \approx 3.1$ instead of the regular value of 2. This enhancement can be understood when considering the complete electronic structure of both the Ag(111) surface state and the Fe atom, as shown by \textit{ab initio} calculations of the magnetic susceptibility.
\end{abstract}

\maketitle

Magnetic excitations of single transition metal atoms adsorbed on a non-magnetic surface, as probed locally by scanning tunneling microscopy (STM), have been the subject of both recent theoretical and experimental studies because of their relevance in fundamental questions of atomic-scale magnetism and future spin-based technological applications. When the transition metal atom is adsorbed on a thin insulator film~\cite{Heinrich2004,Hirjibehedin2007} or on a semiconductor surface~\cite{Khajetoorians2010} its magnetic excitations are well described by model calculations~\cite{Fern'andez-Rossier2009,Lorente2009,Fransson2009,LothNJP2010,Delgado2011} where the atomic magnetic moment is well derived from the number of unpaired spins residing in the free-atomic like $d$-orbitals of the adsorbate.
However, when the transition metal atom is adsorbed on a metal substrate, spin-orbit coupling and hybridization~\cite{Gambardella2002} with the substrate electrons drastically change its magnetic moment~\cite{Gambardella2003} and its magnetization dynamics~\cite{Muniz2003}.

Indeed, previous STM-based studies of Fe adatoms on Pt(111)~\cite{Balashov2009} and Cu(111)~\cite{KhajetooriansPRL2010} set forth the first experimental evidence of the very strong damping of the magnetic excitation of the atom provided by coupling to Stoner modes (spin-flip scattering) of the metallic substrate upon which it is adsorbed. An excellent account of the data in Ref.~\onlinecite{KhajetooriansPRL2010} was provided by calculations that employ a formalism recently developed ~\cite{Muniz2003,Lounis2010,Costa2010,Lounis2011}. These calculations also predicted that the $g$-factor, which characterizes the Zeeman shift when the excitation is monitored by a local probe, deviates from $g = 2$ by $\approx 10$\%~\cite{Mills1967,Muniz2003,Lounis2010,Lounis2011}.

Here we demonstrate that the $g$-factor of an Fe adatom adsorbed on Ag(111) assumes an anomalously large value exemplifying the strong effect of substrate electrons on the precession of the magnetic moment. Utilizing inelastic scanning tunneling spectroscopy (ISTS), we measure the field-dependent magnetic excitation energy of the Fe atom. The slope of the excitation energy exhibits a $g$-factor of $g = 3.1$, i.e. an enhancement of more than 50\% as compared to the free Fe atom. Moreover, we show that there is a strong anisotropy gap of $\unit[2.7]{meV}$, and a large field-dependent linewidth broadening. We also present calculations based on the formalism we have developed~\cite{KhajetooriansPRL2010,Lounis2010,Lounis2011}, utilizing the Korringa-Kohn-Rostoker Green function (KKR) method~\cite{Papanikolaou2002} within time-dependent density functional theory (TD-DFT)~\cite{Runge1984,Gross1985}, which provide an excellent account of the anomalous $g$-factor found experimentally.  Theoretically, these results are compared to Mn, Cr, and Co atoms in various lattice configurations.


All measurements were obtained in a home-built STM facility with a base temperature of $\unit[0.3]{K}$ and a magnetic field of up to $12$T perpendicular to the sample surface~\cite{Wiebe2004a}. The STM tip was made from high purity Ag wire by cutting and subsequent annealing after transfer into the ultra high vacuum facility. The sample was prepared in situ by repeated cycles of sputtering with Ar$^+$ ions and subsequent annealing at a temperature of $\unit[550]{^{\circ}{C}}$, yielding an atomically flat Ag(111) surface with terraces several 100 nm wide. Single Fe atoms were deposited onto the cold sample kept at a temperature of $T<\unit[6]{K}$. Differential conductance ($\text{d}I/\text{d}V$) spectra were obtained by stabilizing the STM tip above the adsorbate at tunneling current $I_{\rm stab}$ and bias voltage $V_{\rm stab}$ applied to the sample, switching off the feedback circuit and ramping the bias voltage $V$ while recording the $\text{d}I/\text{d}V(V)$ signal via lock-in technique with modulation voltage $V_{\rm mod}=\unit[50-100]{\mu V}$ and modulation frequency $f_{\rm mod}=\unit[4.1]{kHz}$.

\begin{figure}
\includegraphics[width=0.7\columnwidth]{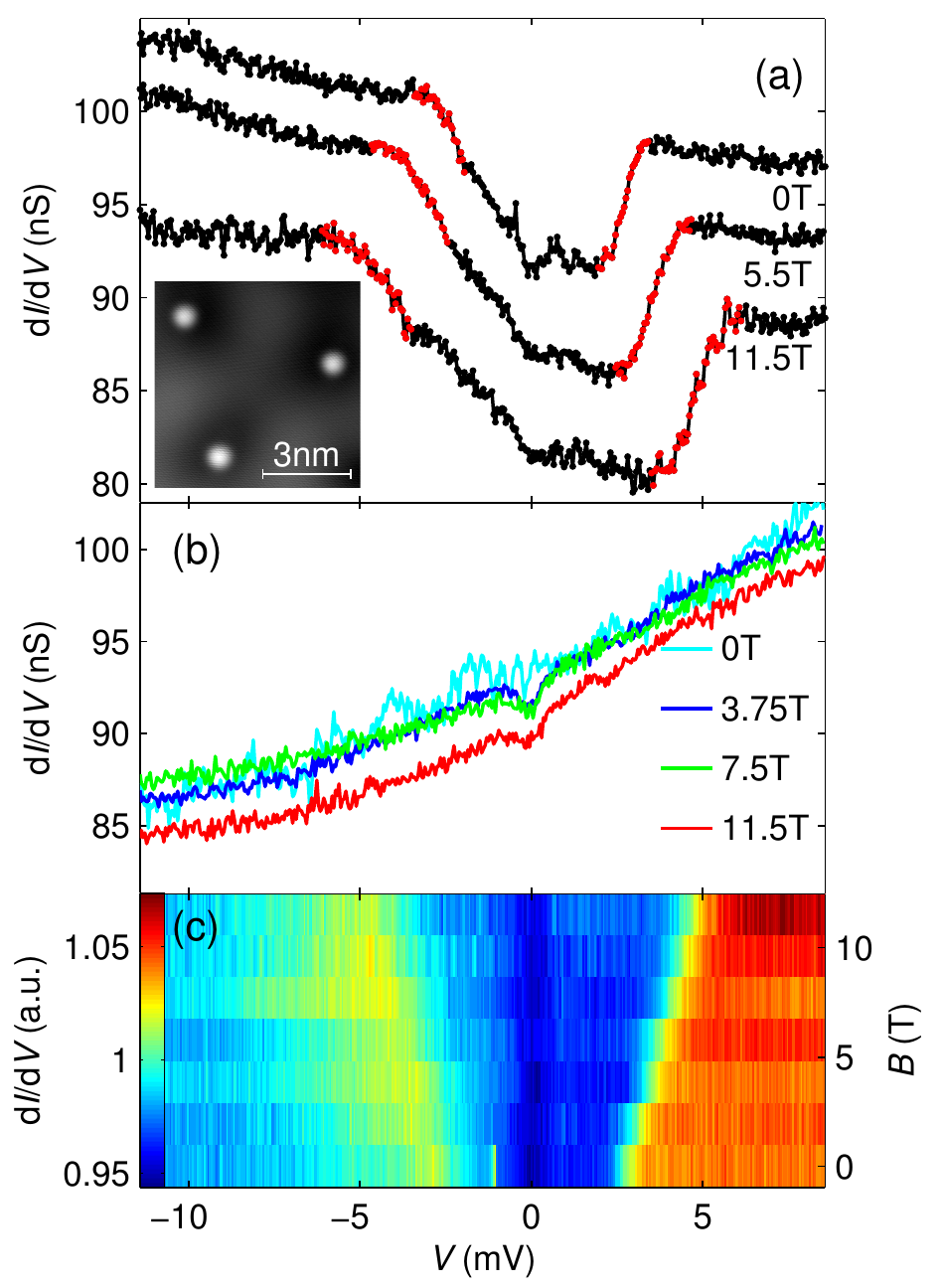}
\caption{(a) ISTS spectra taken above the Fe adatom for different magnetic fields $B$ with the same microtip (spectra at $\unit[5.5]{T}$ and at $\unit[11.5]{T}$ are vertically offset by $\unit[-7]{nS}$ and by $\unit[-12]{nS}$ for clarity). On the positive bias side, the part of the spectrum in the interval $E_{\rm{step}}-w<eV<E_{\rm{step}}+w$ is highlighted in red, where $E_{\rm{step}}$ is the fitted step energy and $w$ is the fitted step width (FWHM, see text). On the negative bias side, the part of the spectrum on the negative of this interval is highlighted. Inset: constant current image of three Fe atoms on Ag(111) ($I_{\rm stab}=\unit[1.3]{nA}$, $V_{\rm stab}=\unit[10]{mV}$, gray scale from $0$ to $\unit[1.35]{\text{\AA}}$). (b) ISTS spectra taken with the same microtip on the substrate. (c) Color scale representation of a similar dataset taken at smaller increments in the magnetic field. Individual spectra were divided by their mean value for better visualization. ($I_{\rm stab}=\unit[1]{nA}$, $V_{\rm stab}=\unit[10]{mV}$, $V_{\rm mod}=\unit[100]{\mu V}$)}
\label{fig:AgISTS}
\end{figure}

Figure~\ref{fig:AgISTS}(a) shows high energy-resolution $\text{d}I/\text{d}V$ spectra taken on an isolated Fe atom (inset) with the same microtip for different magnetic field values. In comparison to the spectra taken on a nearby substrate location (Fig.~\ref{fig:AgISTS}(b)), which were always featureless and almost flat, a reduction in the differential conductance signal around the Fermi energy is present. On the positive bias side, this reduction occurs in the form of a sharp step, whereas on the negative side the decrease of the signal is more gradual. The step on the positive bias side clearly shifts away from the Fermi energy $E_{\rm F}$ ($\unit[0]{V}$) for increasing magnetic field. To further illustrate the evolution of the two spectroscopic features in a magnetic field, Fig.~\ref{fig:AgISTS}(c) shows an intensity plot of the $\text{d}I/\text{d}V$ spectra of a similar dataset, as a function of the magnetic field $B$ which has been incremented in smaller steps. In the intensity plot, the step at positive bias and a step occurring at negative bias symmetric with respect to $E_{\rm F}$ are visible, and linearly shifting toward higher absolute bias voltages for increasing magnetic field. We conclude, that the two steps are due to a magnetic excitation of the Fe atom by inelastic spin-flip scattering of the tunneling electrons~\cite{KhajetooriansPRL2010}. The corresponding steps in the $\text{d}I/\text{d}V$ curves are marked in red in Fig.~\ref{fig:AgISTS}(a). While the excitation step occurs on a rather flat background $\text{d}I/\text{d}V$ signal for positive sample bias, it is superimposed on the gradual increase of the $\text{d}I/\text{d}V$ signal on the negative sample bias side. The gradual increase does not change with the magnetic field and is observed for all atoms in the filled states regime, independent of the utilized microtip. Thus, it is probably due to a peak in the local electron density of states (LDOS) above the Fe atom. In all further analysis, only the unmasked step on the positive bias side was considered.



%

To extract both the $g$-factor and the lifetime of the magnetic excitation, ISTS spectra of $17$ Fe atoms were analyzed for different magnetic fields, giving a total of $70$ datapoints for each of the following quantities: The step position $E_{\rm{step}}$, the step width $w$ (FWHM), and the inelastic contribution to the total $\text{d}I/\text{d}V$ signal for bias voltages greater than the excitation energy, $P_{\rm inel}$. All values have been extracted by fitting a gaussian broadened step function to the measured spectra in the positive bias regime. It turns out that $P_{\rm inel}=\unit[8.8]\% \pm \unit[1.8]\%$ (std) of the tunneling electrons are inelastically scattered at the Fe atom and induce a spin excitation. Therefore, the process is considerably more efficient than for the case of Fe on Cu(111)~\cite{KhajetooriansPRL2010} or Fe on Pt(111)\cite{Balashov2009}.

\begin{figure}[t]
\includegraphics[width=\columnwidth]{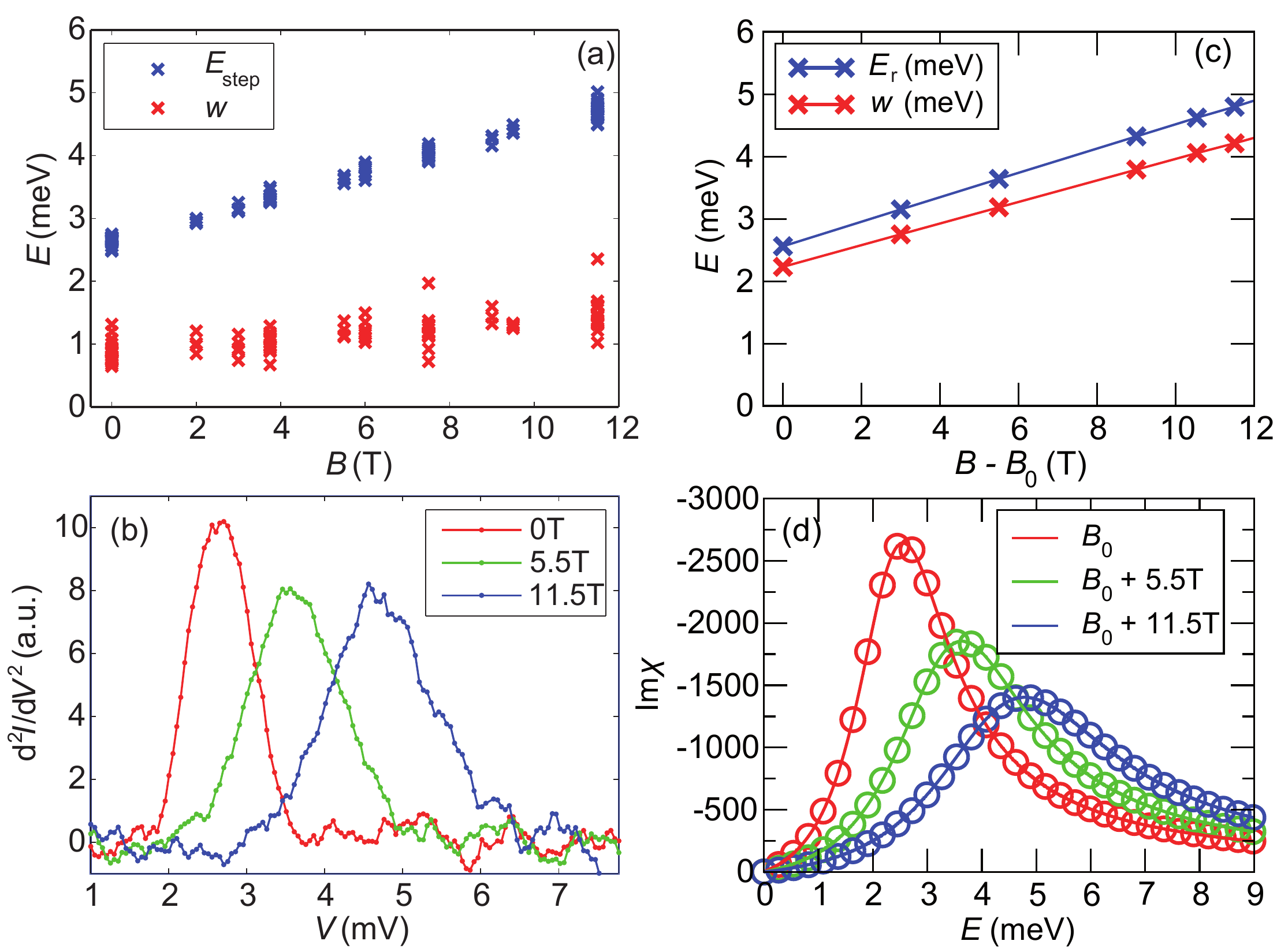}
\caption{(a) Fitted step energy $E_{\rm{step}}$ and step width $w$ (FWHM) of the inelastic step on the positive voltage side of the ISTS spectra taken on 17 different Fe atoms as a function of magnetic field $B$.  (b) Numerical derivative of the positive bias part of the spectra used in Fig. 1(a).  (c) Calculated magnetic field dependency of the resonance energy $E_\mathrm{r}$ and
of the linewidth $w$. (d) Examples of computed density of magnetization excitations for
an Fe adatom on Ag(111).}
\label{fig:Ag:E_vs_B}
\end{figure}

The step energy $E_{\rm{step}}$ and width $w$ are shown in Fig.~\ref{fig:Ag:E_vs_B}(a) as a function of the magnetic field $B$. The step energy shows a zero field splitting of $E_{\rm{step}}(B=\unit[0]{T})=\unit[2.7]{meV}\pm\unit[0.06]{meV}$ (std) and then increases linearly with $B$ with a slope of $3.13 \pm 0.07 \mu_B$ (std). The zero field splitting results from the magnetic anisotropy energy of the Fe moment. If we assume that the tunneling electron spin changes by $\pm 1$ during spin-flip scattering, and assume conservation of the total angular momentum comprised of the sum of the tunneling electron spin and of the total angular momentum of the Fe adatom on Ag(111), the slope of the step energy as a function of the magnetic field results in an effective $g$-factor of the adatom on substrate system of $g = 3.13 \pm 0.07$. Finally, as exemplified by the derivative of the spectra of Fig.~\ref{fig:AgISTS}(a) with respect to the voltage shown in Fig.~\ref{fig:Ag:E_vs_B}(b), there is a significant linewidth broadening of the excitation as the magnetic field is increased. This linewidth increases with a rate of $\approx\unit[1.1]{\mu_B}$.

As we have shown for the case of Fe adatoms on Cu(111)~\cite{KhajetooriansPRL2010}, the linear increase of the linewidth as a function of the excitation energy can be explained by the decay of the excitation into Stoner modes of the itinerant conduction electrons of the substrate~\cite{Muniz2003,Lounis2010,KhajetooriansPRL2010,Lounis2011}.
Due to the hexagonal symmetry of the substrate, the measured magnetic anisotropy energy for Ag(111) most probably favors a uniaxial out-of plane orientation of the Fe moment as predicted with our simulations using KKR which reveal a perpendicular magnetic anisotropy of $\unit[5.6]{meV}$.  Moreover, the experimental value is almost a factor of three larger than for Cu(111)~\cite{KhajetooriansPRL2010}. Also the lifetime of the excitation, which is derived from the linewidth at $B = \unit[0]{T}$ to be $\tau=\hbar/(2\Delta E)\approx\unit[400]{fsec}$ is about twice as large as for Cu(111) and decreases with a rate which is two times smaller. Remarkably, the $g$-factor for the case of Ag(111) is $50$\% larger than for Cu(111).

In order to elucidate the $g$-factor enhancement, first principles calculations have been carried out to
 evaluate the energy-dependent local transverse spin susceptibility $\chi(E)$ whose imaginary part ($\mathrm{Im}(\chi)$) describes the density of states of possible magnetic excitations. The formal expression for the dynamic susceptibility we study can be written as
\begin{equation}
\chi = \chi_0 /(1 -U\chi_0),
\label{eq:th1}
\end{equation}
 connecting the Kohn-Sham susceptibility $\chi_0$ (that describes Stoner excitations) to the exchange and correlation kernel $U$.
 To mimic spin-orbit coupling, which is not incorporated into our KKR-based formalism, an appropriate external magnetic field is applied ($B_0$). This approach
produced results in excellent accord with the ETB description of magnetic excitations of the Fe adatom on Cu(111) where spin-orbit coupling was incorporated~\cite{KhajetooriansPRL2010}.

In Fig.~\ref{fig:Ag:E_vs_B}(c) we plot the magnetic-field dependence of the theoretical resonance energy $E_\mathrm{r}$ and linewidth $w$ which can be directly compared to Fig.~\ref{fig:Ag:E_vs_B}(a). We find a $g$-factor of 3.34 which is
in a striking agreement with the measured value. The theoretical linewidth ($B = 0$) is about twice the corresponding experimental value and broadens more heavily ($\approx 2.9\mu_B$) as compared to the averaged measured linewidth ($\approx 1.1\mu_B$). Interestingly, our calculations identify the Fe-adatom $g$-factor enhancement as a rare case, since Cr, Mn and Co adatoms are characterized by typical $g$ values of 2.1, 1.9 and 1.84 (not shown).
%

\begin{figure}[t]
\includegraphics[width=\columnwidth]{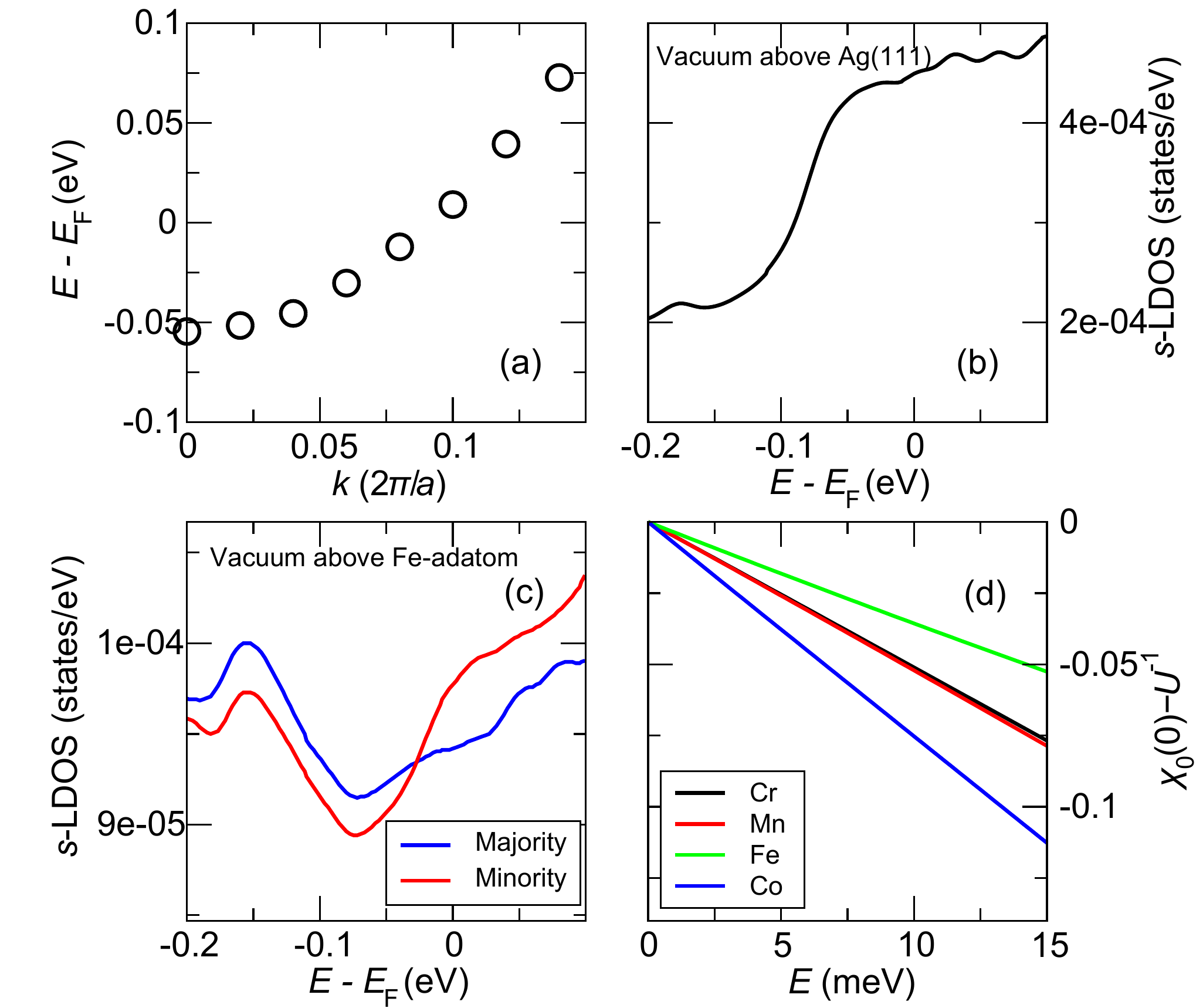}
\caption{(a) Calculated band structure showing the surface state of Ag(111) centered
around the $\Gamma$-point, where $a$ is the lattice parameter of Ag. (b) Corresponding calculated step-like $s$-LDOS in the vacuum evaluated at 4.7~\AA ~ above the surface. (c) If an Fe atom is put on top of the surface, a bound state is created that is spin-splitted. (d) Linear behavior of the real part of the Kohn-Sham susceptibility for different adatoms on Ag(111). The Fe adatom is characterized by the smallest slope explaining its large $g$-shift.}\label{fig_theory3}
\end{figure}

The Ag(111) surface is peculiar since its surface state onset is located much closer to $E_\mathrm{F}$ compared to Cu(111) and Au(111)~\cite{Reinert2001}. Our calculations indicate an onset value of $\unit[50]{meV}$ (Fig.\ref{fig_theory3}(a)-(b)). Moreover, it is well established that an adatom induces a bound state that is split-off from the bottom of the surface state band if the adatom potential is attractive~\cite{Olson2004, Limot2005, Lounis2006}. These effects could principally affect the $g$-factor.  As on the Cu(111) surface, Fe (Fig.\ref{fig_theory3}(c)), as well as Cr, Mn, and Co adatoms (not shown) exhibit a spin-polarized split-off state on the Ag(111) surface. If this bound state is responsible for the large $g$-shift, it should, consequently, also induce a large $g$-shift for Cr, Mn and Co adatoms. However, this is not seen computationally. Furthermore, these impurities were considered as in-atoms (surface layer) and bulk atoms, where no bound-state is present as predicted for the Cu(111) surface~\cite{Lounis2006}. Among all additional impurities investigated, only the Mn in-atom exhibits a large $g$-factor of 2.9 illustrating that the bound state is not necessarily responsible for enhancing the $g$-factor.

Finally, to investigate the influence of the surface state, we shifted its location by reducing the thickness of the simulated slab thickness from 24 to 5 monolayers. This artificial consideration modifies the surface state since the two surface states at both ends strongly interact thereby splitting into bonding and anti-bonding states. These are then located very far from $E_\mathrm{F}$. This electronic modification diminishes the $g$-factor of the Fe adatom to $g \approx 2$.  This indicates that the electronic properties of the surface state play an important role in the observed $g$-shift.

While there is not a simple picture of how the interplay of the surface state electronic structure combined with that of the Fe atoms produces the observed $g$-shift, a specific trend can be seen in the low energy properties of the susceptibility which may explain the observed $g$-shift.  Since we probe energies far below the
electronic scale, we can expand $\chi_0(E)$ in powers of the
energy $E$ if desired.  For small energies, it can be shown that the real part $\mathrm{Re}(\chi_0)$ and the imaginary part $\mathrm{Im}(\chi_0)$ of the Kohn-Sham susceptibility are linear functions of $E$,
{\it i.e.} $\mathrm{Re}(\chi_0 (E)) = \chi_0(0) + \alpha E $ and
$\mathrm{Im}(\chi_0 (E)) = \beta E$. $\chi_0(0)$ is nothing else than the static susceptibility and $\mathrm{Im}(\chi_0)$
describes the density of Stoner modes, while $\alpha$ and
$\beta$ are the slopes defining the linear energy dependence. Analytically, $\beta$ is given by the product between the spin-dependent adatom density of states at $E_\mathrm{F}$ ($-\pi n_{\downarrow}(E_\mathrm{F})n_{\uparrow}(E_\mathrm{F})$), while $\alpha$ is more complex but can be expressed in terms of single particle
Green functions evaluated at $E_\mathrm{F}$.

In Fig.~\ref{fig_theory3}(d), $\mathrm{Re}(\chi_0)$ at $B = 0$ is plotted against energy for Cr, Mn, Fe and Co adatoms on Ag(111) and indeed the trend is linear and is not modified with an applied field. Interestingly, $\alpha$  is smallest for the Fe-adatom, which strongly contributes to the large $g$-shift. In fact, by plugging the linear behavior of $\chi_0$ into Eq.~\ref{eq:th1} and evaluating $\mathrm{Im}(\chi)$ we obtain an equation similar to what is given in~\cite{Mills1967,Muniz2003}:
\begin{equation}
\mathrm{Im}(\chi) = \frac{\beta E}{(1-U(\chi_0(0) + \alpha E ))^2 + (U\beta E)^2}\label{eq:th2}
\end{equation}
where the resonance energy $E_\mathrm{r} = \frac{|\frac{1}{U}- \chi_{0}(0)|}{\sqrt{\alpha^2+\beta^2}}$.
If no external magnetic field is applied along the $z$-direction, $\chi_0(0)=1/U$~\cite{Lounis2010,Lounis2011} and
consequently $E_\mathrm{r} = 0$. \emph{This is a key result}: the position of the resonance, and thus the $g$-shift, depends equally on the slope of $\mathrm{Re}(\chi_0)$, the slope of $\mathrm{Im}(\chi_0)$, and the change of $\chi_0$ at zero energy induced by the magnetic field. Thus, the right combination of these properties have to be satisfied in order to observe a large $g$-shift as for the Fe adatom.

This new data illustrates that 3\it{d} \rm{}transition metal moments, when excited and monitored by a local probe, can display a wide range of response characteristics not realized with long wavelength probes, as in ferromagnetic
resonance or Brillouin light scattering spectroscopy. The data also provided the first experimental and \emph{ab-initio} verification of a prediction made many years ago: Coulomb interactions alone lead to strong damping of the spin precession of a single moment, when it is both excited and its motion monitored by a local probe that samples its large wave vector response~\cite{Mills1967}.  By considering the detailed electronic structure, it is possible to realize enhanced $g$-factors of single atoms on a surface which is unimaginable for single atoms well described by an ionic model.

We would like to thank P. Gambardella for fruitful discussions.  We acknowledge funding from SFB668-A1 and GrK1286 of the DFG, from the ERC Advanced Grant ``FURORE", and from the Cluster of Excellence ``Nanospintronics". The research of D. L. M. and A. T. C. was supported by the U. S. Dept. of Energy, via grant DE-FG03-84ER-45083. S. L. acknowledges the support of the HGF-YIG Programme VH-NG-717.
%
%


\end{document}